\documentclass{article}
\usepackage[utf8]{inputenc}
\usepackage[english]{babel}
\usepackage{hyperref}
\usepackage{enumitem}
\usepackage{relsize}
\usepackage{xcolor}
\usepackage{eufrak}
\usepackage{amsmath}
\usepackage[sort]{cite}
\usepackage{mathrsfs}
\usepackage[left=2.5cm,right=2.5cm,top=2.5cm,bottom=2.6cm]{geometry}
\usepackage{multicol}

\newcommand{\ham}{\mathcal{H}}
\newcommand{\diff}{\mathcal{D}}
\newcommand{\shift}{{N^r}}
\newcommand{\lapse}{N}
\newcommand{\vars}{{E^r},{E^\varphi},{\phi},{K_r},{K_\varphi},{P_\phi}}

\title{\textbf{Anomaly-free deformations of spherical general relativity coupled to matter\\
}}
\author{{{Asier Alonso-Bardaji}\footnote{asier.alonso@ehu.eus}\,} and {{David Brizuela}\footnote{david.brizuela@ehu.eus}}}

\date{}

\newenvironment{changemargin}[2]{%
	\begin{list}{}{%
			\setlength{\topsep}{0pt}%
			\setlength{\leftmargin}{#1}%
			\setlength{\rightmargin}{#2}%
			\setlength{\listparindent}{\parindent}%
			\setlength{\itemindent}{\parindent}%
			\setlength{\parsep}{\parskip}%
		}%
		\item[]}{\end{list}} 

\begin{document}

\maketitle

\vspace*{-1cm}

\begin{center}
    {\emph{\small{Fisika Saila, Universidad del Pa\'is Vasco/Euskal Herriko Unibertsitatea (UPV/EHU),\\Barrio Sarriena s/n, 48940 Leioa, Spain}}}\\
\end{center}
\vspace{8pt}

\begin{changemargin}{1.5cm}{1.5cm}
\begin{center}
    \textbf{Abstract}
\end{center}
\vspace{4pt}
\noindent A systematic approach is developed in order to obtain spherically symmetric midisuperspace
models that accept holonomy modifications in the presence of matter fields with local degrees of freedom.
In particular, starting from the most general Hamiltonian quadratic in radial derivatives of the variables,
we obtain a family of effective modified constraints that satisfy Dirac's deformation algebra,
which encodes the covariance of general relativity, and show that (scale-dependent) holonomy corrections
can be consistently implemented.
In vacuum, the deformed anomaly-free Hamiltonian is explicitly written in terms of three free functions 
and we obtain a weak observable that can be interpreted as the mass of the model.
Finally,
as a particular example, we present a specific covariant polymeric model that remains regular for any
value of the connection components. Some of its physical implications and the relation with previous
studies in the literature are commented.\\
\end{changemargin}
\vspace{8pt}

\section{Introduction}\label{intro}

In the context of effective theories, holonomy-modified cosmological models provide an accurate
description when compared to the full dynamics predicted by the equations of loop quantum cosmology.
The usual approach to obtain such effective descriptions is to modify the Hamiltonian by hand so
that the expected effects from loop quantum gravity are included. In loop quantum cosmology, holonomy corrections are directly related to the spacetime discreteness and they are shown to solve the initial singularity.
Hence, in this paper, we will focus on this kind of modifications in our study of effective spherically
symmetric midisuperspaces.

Homogeneous configurations present an off-shell vanishing diffeomorphism constraint and one can include
by hand a wide variety of modifications in the Hamiltonian, as the closure of the Poisson algebra
is trivially guaranteed. The initial cosmological studies yielding a quantum bounce at early times
for isotropic models
were rapidly extended to anisotropic Kantowski-Sachs spaces. These models describe the homogeneous
interior region of black holes \cite{AshtekarBojowald,gambini&pullin,corichibh,Barrau:2018rts,pertashtekar,Ashtekarkruskal,tecotl,bmm,Bouhmadi2020,olmedobh} and showed interesting predictions. For instance, the implementation of holonomy corrections points towards a transition from black to white holes.

However, obtaining a consistent theory becomes rather challenging when non-homogeneous scenarios (particularly those with local degrees of freedom) are considered. In these cases, the diffeomorphism constraint is no longer vanishing and one finds that the quantized notion of spacetime collides with the continuous diffeomorphism symmetry of general relativity. To circumvent these difficulties, several approaches have been presented in the literature. For instance,
the hybrid quantization program \cite{MartinBenito:2008ej,FernandezMendez:2012vi,Fernandez-Mendez:2013jqa,1407.0998,PhysRevD.93.104025} considers a homogeneous background spacetime
quantized via loop-quantum-gravity techniques, and incorporates inhomogeneities by means of a Fock quantization. Similar assumptions are made in the dressed-metric approach \cite{PhysRevD.87.043507,Agullo_2013, Agullo_2017}, providing a completely quantized theory to study cosmological perturbations. However, it remains unclear whether these frameworks respect general covariance \cite{PhysRevD.102.023532}.

Spherically symmetric configurations are of particular relevance as they would provide a first step towards
the description of black holes and gravitational collapse. It was found that the Abelianization of constraints could lead to a consistent quantization even under the presence of a scalar matter field \cite{PhysRevLett.110.211301,Gambini_2014}. These studies have been extended to include a possible polymerization of the matter component \cite{bgp1,bgp2,bgp3},
so that holonomy effects can be explored even in the radial gauge with the polar foliation condition. However, recent discussions cast doubts on the covariance of these models
\cite{bojo2,Bojowald:2021isp}. {  In addition, one should also mention current works where,
as in the improved dynamics scheme of loop quantum cosmology, scale-dependent holonomies are implemented in vacuum spherical symmetry \cite{Kelly:2020uwj,Gambini:2020qhx,Han:2019feb}.
Nevertheless, these models break covariance when they are weakly coupled to a matter field without further modifications.}

In this context, we find the consistent constraint deformation formalism, which searches for effective theories that explicitly satisfy covariance through Dirac's deformation algebra. This methodology has been successfully implemented in the study of cosmological perturbations \cite{anomalpert,pert1,pert2,Cailleteau:2011kr,pertnew2,pertholonomy,Wilson-Ewing:2011gnq} and spherically symmetric models \cite{tibrewala,algebroids,Bojowald:2018xxu,nodiff}. Particularly, it is argued that the effective line element corresponding to the holonomy-corrected constraints might present a signature change when approaching the classical singularity \cite{Bojowald:2018xxu}. From this perspective, some interior region of the black hole would be Euclidean, and the usual notions of causality and time evolution would not apply. Unfortunately, strong no-go results point out that these models can not be extended to include matter fields with local degrees of freedom \cite{Alonso-Bardaji:2020rxb}, {  although it may be possible to evade such no-go results by using self-dual variables }\cite{BenAchour:2016brs,BenAchour:2017jof,Bojowald:2019fkv}. 
This could mean that the possibility of polymerizing the curvature components is a consequence of an excessive symmetry assumption. Indeed, homogeneous models in effective loop quantum cosmology accept any kind of polymerization for the connection components 
but even the simplest non-homogeneous cases (spherically symmetric vacuum) present limitations regarding polymerization due to the first-class nature of Dirac's deformation algebra. These restrictions are stronger when one adds matter with local degrees of freedom to the system. 

In this paper,
we study a simple midisuperspace model and show that one can in fact consider holonomy corrections (even scale-dependent ones) in the presence of local degrees of freedom.
In Sec.~\ref{classical} we briefly review the classical canonical formulation
of spherically symmetric spacetimes and introduce the dynamics through a scalar matter field.
In Sec.~\ref{modified}, we will show that the requirement of anomaly freedom gives rise to severe consistency conditions, which are necessary for our theory to remain covariant. More precisely, we will compute Dirac's algebra for a generic Hamiltonian quadratic in radial derivatives of the configuration variables along with the classical diffeomorphism constraint. Although straightforward, these computations
are extremely lengthy and, in certain cases,
we will omit some intermediate steps in order to show only the final result.
In Sec.~\ref{vacuum}, we completely solve the vacuum anomaly equations and provide the most general anomaly-free Hamiltonian that is quadratic in radial derivatives of the triad components. Moreover, we find an observable of this system, which we identify with the mass of the model. In Sec.~\ref{polymerization}, we reconsider the case coupled to matter and obtain a specific polymeric constraint that shows a regular behavior for any value of the angular component of the connection. The previous notion of
mass will be extended to this model and the deformation of the algebra will be given in terms of scalar functions. Finally, Sec.~\ref{conclusion} summarizes the results and presents our conclusions.

\section{Classical midisuperspace formulation}\label{classical}

In the canonical formalism of general relativity the total Hamiltonian is a combination of first-class constraints.
These constraints govern the dynamics and, at the same time, encode the covariance of the theory, which is not
explicit in this approach. Rather the commutation relations between constraints, the so-called Dirac's hypersurface algebra, ensure that the theory remains covariant.

Our analysis will be restricted to spherically symmetric models. In this case, there is only one non-homogeneous space direction and thus two components of the diffeomorphism constraint will vanish by choosing coordinates adapted to the symmetry. Instead of the four constraints from the full theory, we retain only two. These are the Hamiltonian
${\cal H}^{(class)}$ and the radial component of the diffeomorphism constraints ${\cal D}$
which, in a smeared form, will be written as,
\begin{align}
 H^{(class)}[\lapse]:=\int{dr}\, \lapse\ham^{(class)}\,, \quad\text{and}\quad D[\shift]:=\int{dr}\, \shift\diff.
\end{align}
Here, $r$ denotes the non-homogeneous coordinate, $\lapse$ is the lapse and $\shift$ is the only non-vanishing component of the shift.
In addition, the dependence on the homogeneous coordinates has been integrated out and all global constants have been set to one. The geometric degrees of freedom will be described by the $U(1)$ invariant components of the spherically symmetric triad, $E^r$ and $E^\varphi$, combined with their conjugate
momenta, $K_r$ and $K_\varphi$. The symplectic structure is given by,
\begin{align}
    \{E^r(r_a),K_r(r_b)\}=\{{E^\varphi}(r_a),{K_\varphi}(r_b)\}=-\delta(r_a-r_b).
\end{align}
As we want to study dynamical scenarios, we need to couple some matter field to the model. Particularly, we will consider a scalar field $\phi$,
with canonical Poisson brackets with its conjugate momentum $P_\phi$,
\begin{align}
    \{{\phi}(r_a),{P_\phi}(r_b)\}=\delta(r_a-r_b).
\end{align}
In terms of the Ashtekar variables, the classical diffeomorphism constraint takes the form,
\begin{align}
    &\label{diff}\diff =
     -{E^r}'{K_r} +{E^\varphi} K_\varphi' +\phi'P_\phi.
\end{align}
whereas the classical Hamiltonian constraint reads,
\begin{align}\label{hamspherclass}
    \ham^{(\text{class})} &= -\frac{{E^\varphi}}{2\sqrt{{E^r}}} -\frac{{E^\varphi}K_\varphi^2}{2\sqrt{{E^r}}}  -2\sqrt{{E^r}}{K_r}{K_\varphi}   + \frac{({E^r}')^2}{
 8 \sqrt{{E^r}} {E^\varphi}} - \frac{\sqrt{{E^r}}}{2{E^\varphi}^2} {E^r}' {E^\varphi}' +\frac{\sqrt{{E^r}}}{2{E^\varphi}}{E^r}'' +\mathcal{H}_{\mathrm{m}},
\end{align}
where we have considered the positive orientation of the triad. The prime denotes the derivative with respect to $r$ and
$\mathcal{H}_\mathrm{m}=\mathcal{H}_\mathrm{m}({E^r},{E^\varphi},{\phi},{P_\phi})$ is the matter contribution
which, for the minimally coupled scalar field under consideration, is given by:
\begin{align}
    \mathcal{H}_\mathrm{m}=\frac{P_\phi^2}{2\sqrt{{E^r}}{E^\varphi}} +\frac{({E^r})^{3/2}}{2{E^\varphi}}({\phi}')^2 +\sqrt{{E^r}}{E^\varphi}V({\phi}).
\end{align}
In terms of these variables, the spacetime line element can be written as follows,
\begin{align}
    {ds}^2 = -N^2{dt}^2 + \frac{(E^\varphi)^2}{E^r}\left({dr} +N^r{dt}\right)^2 +E^r {d\Omega}^2,
\end{align}
and Dirac's deformation algebra reads,
\begin{subequations}\label{hda}
\begin{align}
    \label{hdadd}\big\lbrace D[\shift_{\!\!1}],D[\shift_{\!\!2}]\big\rbrace &= D\big[\shift_{\!\!1}\shift_{\!\!2}'-\shift_{\!\!1}'\shift_{\!\!2}\big],\\
    \label{hdadh}\big\lbrace D[\shift],H^{(\text{class})}[\lapse]\big\rbrace &= H^{(\text{class})}\big[\shift\lapse'\big],\\
    \label{hhclass}\big\lbrace H^{(\text{class})}[\lapse_1],H^{(\text{class})}\big[\lapse_2\big]\big\rbrace &= D\big[{E^r}({E^\varphi})^{-2}\left(\lapse_1\lapse_2'-\lapse_1'\lapse_2\right)\!\big].
\end{align}
\end{subequations}
Note that the structure function in the last bracket is an element of the inverse metric.

\section{Modified constraint algebra}\label{modified}

The main goal of this paper is to construct the most general Hamiltonian
constraint ${\cal H}$ that is quadratic in radial derivatives of the variables
of our model $(E^r,E^\varphi,\phi)$ and that respects
the covariance of the system. For such a purpose, we begin
with an ansatz of the form,
\begin{align}
    &\label{ham}{\cal H} = 
    a_{0} +({E^r}')^2a_{rr}+({E^\varphi}')^2a_{\varphi\varphi}+({\phi}')^2a_{\phi\phi}+{E^r}'{E^\varphi}'a_{r\varphi}+{E^r}'{\phi}'a_{r\phi}+{E^\varphi}'{\phi}'a_{\varphi\phi} +{E^r}''a_2,
\end{align}
where all quadratic combinations of the {  radial derivatives of $E^r,E^\varphi$ and $\phi$} are included multiplied by a free function $a_{ij}=a_{ij}(E^r,E^\varphi,\phi,K_r,K_\varphi,P_\phi)$, with $i,j=r,\varphi,\phi$,
that depends on all the variables and momenta of the model.
Following the form of the classical Hamiltonian \eqref{hamspherclass}, only one second-order derivative term has been included, that is, 
${E^r}''$, with a free coefficient
$a_2=a_{2}(E^r,E^\varphi,\phi,K_r,K_\varphi,P_\phi)$. {  Other second-order derivatives (${E^\varphi}'',\phi''$) are not expected to affect possible holonomy corrections and would further complicate the algebraic computations. On the other hand,} the free function
$a_0=a_{0}(E^r,E^\varphi,\phi,K_r,K_\varphi,P_\phi)$ includes all the terms that do
not depend on radial derivatives. {  Finally, note that no radial derivatives of the curvature components are considered and, hence, we only expect to be able to describe pointwise holonomy corrections. Further modifications involving derivatives of $K_r$ and $K_\varphi$ are beyond our assumptions.}

In order to fix these free functions, two conditions will be imposed. First,
the above generic expression \eqref{ham} will be required to form an anomaly-free algebra
along with the classical diffeomorphism constraint ${\cal D}$ \eqref{diff}. Second,
the classical limit for the constraint $\ham$ should be given by the classical
Hamiltonian ${\cal H}^{(class)}$ \eqref{hamspherclass}.
As we will see,
these two requirements severely restrict the form of the free functions in ${\cal H}$. 

As a side note, since the density weight of the different objects will be
of key relevance in the following discussion, let us
recall that the Hamiltonian \eqref{ham} must be a weight-one density.
Moreover, from each conjugate couple, one of them is a scalar whereas the other one
is a weight-one density. In fact, the diffeomorphism constraint \eqref{diff} shows explicitly
this weight distribution: those variables that appear primed in \eqref{diff}, that is (${E^r}$, ${K_\varphi}$, ${\phi}$), are scalars  while their conjugates
(${K_r}$, ${E^\varphi}$, $P^{\phi}$) are densities. To reach that conclusion,
one only needs to identify the gauge transformation generated by the constraint $D[\shift]$
on each canonical variable
with its Lie derivative along the shift vector.

\subsection{$\mathbf{\{D,H\}}$ bracket}\label{dhbracket}

The Poisson bracket between the classical diffeomorphism \eqref{diff}
and the modified Hamiltonian constraint \eqref{ham} is quite lengthy and contains
several anomalous terms. In order to classify and solve these anomalies
systematically, we write the result in
a unique way by removing all the derivatives from the shift vector
through integration by parts. In this way, the result
can be schematically written as,
\begin{align}\label{dhbracketf}
    \big\lbrace D[\shift],H[\lapse]\big\rbrace = \int{dr} \,\shift\left(\lapse\mathcal{F}_0 +\lapse'\mathcal{F}_1 +\lapse''\mathcal{F}_2\right),
\end{align}
where $H[N]:=\int dr N {\cal H}$ denotes the smeared form of ${\cal H}$;
and ${\cal F}_0$, ${\cal F}_1$ and ${\cal F}_2$ are complicated expressions
of the free functions $a_{ij}$, $a_0$ and $a_2$, the basic variables of our model and their radial derivatives.
For instance, the term ${\cal F}_2$ is given by
\begin{align}\label{eq.ano2}
    \mathcal{F}_2 =& \left(a_2 +{E^\varphi} a_{r\varphi} \right){E^r}' +2{E^\varphi} a_{\varphi\varphi} {E^\varphi}' +{E^\varphi} a_{\varphi\phi} {\phi}'.
\end{align}
{  Since the free functions $a_0$, $a_{ij}$ and $a_2$ do not depend on radial derivatives,
${\cal F}_2$ can not be written as a linear combination of the constraints ($\diff$ and $\ham$)
and it thus must be regarded as anomalous.} The requirement ${\cal F}_2=0$ is then translated to three independent
equations,
\vspace{-12pt}
\begin{multicols}{3}
\begin{align}
    a_2+{E^\varphi} a_{r\varphi}&=0 ,\label{eq.a2}
    \end{align}
    \columnbreak
    \begin{align}
    &\nonumber\\
    a_{\varphi\varphi}&=0,\label{eq.a22}
    \end{align}
        \columnbreak
    \begin{align}
        &\nonumber\\
    a_{\varphi\phi}&=0\label{eq.a23},
\end{align}
\end{multicols}
\vspace{-12pt}
\noindent which can readily be solved for $a_2$, $a_{\varphi\varphi}$ and $a_{\varphi\phi}$.
Once these relations are enforced, one can continue solving
the remaining anomalies. For example, the coefficient of the term $\shift\lapse'{E^r}'{\phi}'$
on the right-hand side of \eqref{dhbracketf} reads,
\begin{align}\label{eq.ano1}
    {E^\varphi}\frac{(a_{r\varphi})^2}{a_2}\Bigg( {K_r}\frac{\partial}{\partial {K_r}} +{E^\varphi}\frac{\partial }{\partial {E^\varphi}} +{P_\phi}\frac{\partial}{\partial {P_\phi}} -1\Bigg) \frac{a_{r\phi}}{a_{r\varphi}}.
\end{align}
This anomaly will vanish if the function $a_{r\phi}$ takes
the form, 
\begin{align}\label{eq.a13}
    a_{r\phi}={E^\varphi}a_{r\varphi} b_{r\phi}\left({E^r},{\phi},{K_r}/{E^\varphi},{K_\varphi},{P_\phi}/{E^\varphi}\right),
\end{align}
for a generic function $b_{r\phi}$.
Note that the weight-one densities $(E^\varphi,K_r,P_\phi)$ appear as ratios in the arguments of
$b_{r\phi}$, ensuring that it behaves as a scalar function. We must remark that, in order to obtain this solution,
{{$a_2$}} has been assumed to be non-vanishing. Otherwise, there would not be any second-order derivative
in the Hamiltonian \eqref{ham} and the basis of our study would change. With this assumption,
$a_{r\varphi}$ can neither be vanishing as can be read from relation \eqref{eq.a2}.

In fact, it is possible to analytically solve all the anomalies that appear in \eqref{dhbracketf},
and the commented dependence
on scalar quantities 
is dragged along for the rest of the free functions.
More explicitly, the modified Hamiltonian constraint that provides a weakly vanishing
bracket with the classical diffeomorphism constraint takes the form,
\begin{align}\label{hamdh}
    {\cal H} &= -\sqrt{{E^r}}\frac{g}{2}\bigg({E^\varphi} b_0 +{E^\varphi}^{-1}({E^r}')^2b_{rr} -{E^\varphi}^{-1}({\phi}')^2b_{\phi\phi} +{E^\varphi}^{-1}{E^r}'{\phi}'b_{r\phi}+{E^\varphi}^{-2}{E^r}'{E^\varphi}'-{E^\varphi}^{-1}{E^r}''\bigg),
\end{align}
where $b_0=b_0\left({E^r},{\phi},{K_r}/{E^\varphi},{K_\varphi},{P_\phi}/{E^\varphi}\right)$ and $b_{ij}=b_{ij}\left({E^r},{\phi},{K_r}/{E^\varphi},{K_\varphi},{P_\phi}/{E^\varphi}\right)$, with $i,j=r,\phi$,
are free functions of all possible scalar combinations of the model.
In addition, we have defined $g(\vars)$ as a generic global factor.
The anomaly-free bracket between this constraint
and the diffeomorphism constraint reads,
\begin{align}
     \big\lbrace D[\shift],{H}[\lapse]\big\rbrace = {H}\big[\shift\lapse'\big]
     +H\big[\mathcal{G}\big],
\end{align}
where the smearing $\mathcal{G}$ is a function of $\lapse$, $\shift$, $g$ and their derivatives. However, for the Hamiltonian to be a density of weight $+1$, the global factor $g$ must be a scalar function,
and therefore all its arguments must behave as scalar quantities: $g=g\left({E^r},{\phi},{K_r}/{E^\varphi},{K_\varphi},{P_\phi}/{E^\varphi}\right)$. Regarding the Poisson structure, this condition imposes
the vanishing of $\mathcal{G}$. Therefore, the constraint \eqref{hamdh}, with any scalar global factor $g$,
provides the classical result \eqref{hdadh} for the bracket with the diffeomorphism constraint. Note again that the term inside brackets in \eqref{hamdh} is a weight-one density and that the arguments of the different free functions are all the
scalar objects that can be constructed within the model.

\subsection{$\mathbf{\{H,H\}}$ bracket}\label{hhbracket}

We continue our study by computing the Poisson bracket of the constraint \eqref{hamdh} with itself.
In order to analyze the result, we define the combination,
\begin{align}
    \mathcal{N}:= \lapse_1\lapse_2'-\lapse_1'\lapse_2.
\end{align}
It can be shown that, by performing a number of integration by parts,
the commented Poisson bracket can be written as,
\begin{align}\label{hhbracketf}
    \big\lbrace H[\lapse_1],H[\lapse_2]\big\rbrace = \int{dr}\,\mathcal{N}\mathcal{F},
\end{align}
where ${\cal F}$ is again a long expression but it does not depend
on the smearing functions $N_1$ and $N_2$. Just as in the previous section, each coefficient in front of
a given radial derivative must vanish {  on-shell} by itself. In particular,
several of these coefficients
involve partial derivatives of the free functions $b_0$ and $b_{ij}$ with respect to $K_r$ and $P_\phi$, and the requirement of anomaly freedom immediately
fixes the dependence of the Hamiltonian on all the densities $(K_r,E^\varphi,P_\phi)$, except in the global factor $g$.
Hence, let us define the new modified constraint:
\begin{align}\label{hamalfa0}
    {\cal H} &= -\sqrt{{E^r}}\frac{g}{2}\Bigg({E^\varphi} \left(f_{0} +\frac{{K_r}}{{E^\varphi}} f_{1}+\frac{{P_\phi}}{{E^\varphi}}h_{0} +\frac{{P_\phi^2}}{{E^\varphi}^2}f_{3} \right)
    +\frac{({E^r}')^2}{{E^\varphi}}\left(f_{2} +\frac{{K_r}}{{E^\varphi}} h+\frac{{P_\phi}}{{E^\varphi}}h_{1} +\frac{{P_\phi^2}}{{E^\varphi}^2}h_{2} \right)\nonumber\\
    &+\frac{({\phi}')^2}{{E^\varphi}} f_{4} +\frac{{E^r}'{\phi}'}{{E^\varphi}}\left(h_{3} +\frac{{P_\phi}}{{E^\varphi}}h_{4}\right) +\frac{{E^r}'{E^\varphi}'}{{E^\varphi}^2}-\frac{{E^r}''}{{E^\varphi}}\Bigg),
\end{align}
where $h=h({E^r},{\phi},{K_\varphi})$, $f_i=f_i({E^r},{\phi},{K_\varphi})$ and $h_i=h_i({E^r},{\phi},{K_\varphi})$, with $i=0,1,2,3,4$, are free functions of the scalar objects $({E^r},{\phi},{K_\varphi})$.
In addition,
all the functions $h_i$ and $h$ must disappear in the classical limit, whereas the classical expression \eqref{hamspherclass} for the Hamiltonian
is obtained for the following values,
\begin{equation}\label{classlimit}
 f_0\rightarrow \frac{1}{E^r}(1+K^2_\varphi) ,\quad f_1\rightarrow 4K_\varphi,\quad
 f_2\rightarrow-\frac{1}{4E^r},\quad 
 f_3\rightarrow-\frac{1}{E^r},\quad
 f_4\rightarrow-E^r,\quad \text{and} \quad
 g\rightarrow 1.
\end{equation}
At this point, the fixed dependence of the above Hamiltonian
on the densitized variables demands that each coefficient that multiplies a given power of the form
$({K_r})^n({E^\varphi})^m({P_\phi})^l$ on the right-hand side
of \eqref{hhbracketf} must vanish by itself.
In this way, it is possible to rewrite the requirement of anomaly freedom of the bracket \eqref{hhbracketf}
as the following system of differential equations:
\vspace{-12pt}
\begin{multicols}{2}
\begin{subequations}\label{consistencyeqs}
\begin{align}
\frac{\partial f_{0}}{\partial {K_\varphi}}&=2 f_{0} h -2 f_{1} f_{2}-h_{0} h_{3}+\frac{\partial f_{1}}{\partial {E^r}},\label{eq.ed1}\\\label{eq.ed2}
\frac{\partial f_{1}}{\partial {K_\varphi}}&=f_{1} h_{4}+4 f_{3} f_{4},\\
\frac{\partial f_{2}}{\partial {K_\varphi}}&=-h_{1} h_{3}+\frac{\partial h}{\partial {E^r}},\label{eq.ed4}\\
\frac{\partial f_{4}}{\partial {K_\varphi}}&=2 f_{4} (h-h_{4}),\\
\frac{\partial f_{3}}{\partial {K_\varphi}}&=-2 f_{1} h_{2}+2f_{3}(h -h_{4}),\\
\frac{\partial f_{1}}{\partial {\phi}}&=f_{1} h_{3}+2 h_{0} f_{4},\label{eq.ednoKphi}
\end{align}
\columnbreak
\begin{align}
&\nonumber\\
 \frac{\partial h_{0}}{\partial {K_\varphi}}&=-2 f_{1} h_{1}+ h_{0}(2 h-h_{4})-2 f_{3} h_{3},\\
\frac{\partial h_{1}}{\partial {K_\varphi}}&=-h_{1} h_{4}-2 h_{2} h_{3},\\
\label{eq.f32}\frac{\partial h_{2}}{\partial {K_\varphi}}&=-2 h_{2} h_{4},\\
\frac{\partial h_{3}}{\partial {K_\varphi}}&=h_{3}(h - h_{4})-2 h_{1} f_{4}+\frac{\partial h}{\partial {\phi}},\\
\frac{\partial h_{4}}{\partial {K_\varphi}}&=h_{4}(h -h_{4}) -4 h_{2} f_{4}+\frac{\partial h}{\partial {K_\varphi}}.
\label{eq.ed11}
\end{align}
\end{subequations}
\end{multicols}
\vspace{-12pt}
\noindent
This system is composed by eleven first-order partial differential equations
for the eleven free functions $f_i$, $h_i$, and $h$.
Note in particular that the global factor $g$ does not enter the anomaly resolution. 
Even if this system is linear in derivatives, the quadratic combinations of functions
make it very difficult to provide a general solution. Nonetheless, since all the
equations, except \eqref{eq.ednoKphi}, contain derivatives
with respect to $K_\varphi$, it is clear that the dependence on this variable
will be severely restricted. In particular, in Sec.~\ref{polymerization}, we will solve this
system for possible polymeric deformations of the Hamiltonian and, under
some specific choices, we will be able to keep just one free function of $K_\varphi$
in the Hamiltonian.

Whenever the above differential relations are satisfied, the constraint \eqref{hamalfa0}
forms a first-class algebra with the classical diffeomorphism constraint and 
it weakly commutes with itself:
{$\big\lbrace H[\lapse_1],H[\lapse_2]\big\rbrace \approx 0$ }.
The result of this last bracket is rather complicated. Nonetheless, it can be simplified by choosing a global factor that only depends on the scalar quantities, i.e. $g=g({E^r},{\phi},{K_\varphi})$. In this case,
the off-shell bracket reads,
\begin{align}\label{cc}
   \big\lbrace H[\lapse_1],H[\lapse_2]\big\rbrace &= D\left[\mathcal{N}\frac{{E^r}}{{E^\varphi}^2}\frac{g^2}{4}\left(\frac{\partial f_1}{\partial K_\varphi}+\left(\frac{{E^r}'}{{E^\varphi}}\right)^2\frac{\partial h}{\partial {K_\varphi}}\right)\right]
    -H\left[\mathcal{N}\frac{{E^r}'}{{E^\varphi}^2}\sqrt{E^r}\left(h g+\frac{1}{2}\frac{\partial g}{\partial {K_\varphi}}\right)\right],
\end{align}
where it is immediate to check that both smearing functions are of the appropriate weight.
This result differs from the classical case \eqref{hhclass} by an extra $H$ term on the right-hand side{ , which means that the constraint $\ham$ generates a combination of normal and tangential deformations on the hypersurfaces. In order to recover the canonical form of this bracket,
so that the Hamiltonian constraint becomes the infinitesimal generator of normal transformations,}
one 
only needs to demand 
the vanishing of the expression inside the last
round brackets in \eqref{cc}. This leads
to choose a specific global factor $g$ in terms of the function $h$, that is,
$g=\exp(-2\int h\,  d K_\varphi)$
\footnote{In principle one could take $g=\widetilde{g}(E^r,\phi)\exp(-2\int h\, d K_\varphi)$
and leave a free global factor $\widetilde{g}$ that depends on $E^r$ and $\phi$. Nonetheless,
for simplicity we will fix
$\widetilde{g}(E^r,\phi)=1$.}, and further restricts the allowed
dependence on $K_\varphi$ of the Hamiltonian.
Note that, in particular, this choice of global factor is consistent with the classical
limit \eqref{classlimit}.

In summary, we conclude that the generator of infinitesimal normal transformations is given by
the expression \eqref{hamalfa0}, with the global factor $g=\exp(-2\int h\, d K_\varphi)$.
As long as the free functions $f_i$, $h_i$, and $h$ obey the above consistency equations \eqref{cc},
the Poisson bracket of this constraint with itself will be given by,
\begin{align}\label{hh}
   \big\lbrace H[\lapse_1],H[\lapse_2]\big\rbrace &= D\left[\mathcal{N}\frac{{E^r}}{{E^\varphi}^2}\frac{g^2}{4}\left(\frac{\partial f_1}{\partial K_\varphi}+\left(\frac{{E^r}'}{{E^\varphi}}\right)^2\frac{\partial h}{\partial {K_\varphi}}\right)\right].
\end{align}

\section{Reduction to vacuum}\label{vacuum}

In the absence of matter, the generic Hamiltonian \eqref{hamalfa0} derived in the previous section reads,
\begin{align}
    \label{hamSO3vac}\ham^{\rm (vacuum)} &= -\sqrt{{E^r}}\frac{g}{2}\Bigg[{E^\varphi}f_{0} +{K_r}f_{1} +\frac{({E^r}')^2}{{E^\varphi}}\left(f_{2}+\frac{{K_r}}{{E^\varphi}} h\right)+ \frac{{E^r}'{E^\varphi}'}{{E^\varphi}^2} -\frac{{E^r}''}{{E^\varphi}} \Bigg] ,
\end{align}
where now the free functions $f_i$ and $h$ depend only on ${E^r}$ and ${K_\varphi}$, and the global factor is still given
by the relation $g=\exp{[-2\int h{{\rm d}{K_\varphi}}]}$. These functions are not completely independent since
they must satisfy the following two conditions:
\vspace{-12pt}
\begin{multicols}{2}
\begin{subequations}\label{consistencyeqsvac}
\begin{align}
    \frac{\partial f_{0}}{\partial {K_\varphi}}&= 2f_{0}h-2f_{1}f_{2}+\frac{\partial f_{1}}{\partial {E^r}},
    \end{align}
\columnbreak
\begin{align}
&\nonumber\\
    \frac{\partial f_{2}}{\partial {K_\varphi}}&=\frac{\partial h}{\partial {E^r}} .
\end{align}
\end{subequations}
\end{multicols}
\vspace{-12pt}
\noindent In contrast to the case coupled to matter presented in the previous section,
in vacuum it is possible to explicitly solve these equations. The general
solution can be written in terms of the global factor $g(E^r,K_\varphi)$ and two additional integration functions $f=f(E^r,K_\varphi)$ and $v=v(E^r)$, as follows,
\vspace{-12pt}
\begin{multicols}{2}
\begin{subequations}\label{eq.vacsolution}
\begin{align}
  f_0&=\frac{2}{g}\left(\frac{\partial {f^2}}{\partial E^r}+\left(1+f^2\right)\frac{\partial\ln v}{\partial E^r}\right),\\
  f_1&=\frac{2}{g}\frac{\partial {f^2}}{\partial K_\varphi},
  \end{align}
\columnbreak
  \begin{align}
  &\nonumber\\
  f_2&=-\frac{1}{2}\frac{\partial \ln{(g v)}}{\partial E^r},\\
  h&=-\frac{1}{2}\frac{\partial \ln{g}}{\partial K_\varphi},\label{globalfactoreq}
\end{align}
\end{subequations}
\end{multicols}
\noindent where \eqref{globalfactoreq} comes directly from the definition of the global factor.
In this way, the most general anomaly-free vacuum Hamiltonian with quadratic dependence on first-order radial derivatives takes the form,
\begin{align}
    \label{hamSO3vacmod}\ham = -\sqrt{E^r}\Bigg[{E^\varphi}\left(\frac{1+f^2}{v}\frac{\partial{{v}}}{\partial E^r}+\frac{\partial f^2}{\partial E^r}\right) +{K_r}\frac{\partial f^2}{\partial K_\varphi} -g\frac{{E^r}''}{2{E^\varphi}}
    +g\frac{{E^r}'{E^\varphi}'}{2{E^\varphi}^2}-\left(\frac{{E^r}'}{2E^\varphi}\right)^{\!2}\!\left(\frac{E^\varphi}{v}\frac{\partial{{(gv)}}}{\partial E^r}+{K_r}\frac{\partial{{g}}}{\partial K_\varphi}\right)  \Bigg] ,
\end{align}
for any $f(E^r,K_\varphi)$, $g(E^r,K_\varphi)$ and $v(E^r)$ arbitrary functions.
The classical constraint is directly obtained for the values $g=1$, $f= K_\varphi$
and $v=\sqrt{E^r}$. This Hamiltonian constraint generalizes previous models in the literature. For instance, the usual scale-independent polymerization \cite{tibrewala} can be recovered for $g=1$, $v=\sqrt{E^r}$ and $f=\lambda^{-1}\sin(\lambda K_\varphi)$; while the scale-dependent holonomy corrections in \cite{Kelly:2020uwj} correspond to the choice $g=1$, $v=\sqrt{E^r}$ and $f= \sqrt{E^r/\Delta} \sin(\sqrt{\Delta/E^r}K_\varphi)$ along with the partial gauge fixing $K_r=E^\varphi K_\varphi'/(E^r)'$ and $E^r=r^2$.

In addition, it is a straightforward computation to check that the following expression,
\begin{align}\label{mass}
    m = \frac{v}{2}\left(1+f^2-g\left(\frac{{E^r}'}{2E^\varphi}\right)^{\!2}\right),
\end{align}
commutes on-shell with the total Hamiltonian. That is, $\dot{m}=\{m,H[N]+D[N^r]\}\approx 0$, and thus $m$
is a weak Dirac observable of the system. Moreover, its classical limit takes the form,
\begin{align}
    m \rightarrow\frac{\sqrt{E^r}}{2}\left(1+K_\varphi^2-\left(\frac{{E^r}'}{2E^\varphi}\right)^{\!2}\right),
\end{align}
which is the expression of the Schwarzschild mass. In order to check this statement,
one can compute the classical Hawking mass for a spherically symmetric configuration:
\begin{align}\label{massHawking}
    M_H &= \frac{\sqrt{{{E^r}}}}{2}\Big(1-\frac{1}{4{{E^r}}}\partial^\mu{{E^r}}\partial_\mu{{E^r}}\Big) 
    = \frac{\sqrt{{{E^r}}}}{2} \left( 1 +K_\varphi^2 -\left(\frac{{E^r}'}{2{E^\varphi}}\right)^{\!2}\right),
\end{align}
where in the last step the classical evolution equation for $E^r$ has been used. Therefore,
we have obtained an observable \eqref{mass} that can be interpreted as the mass of the deformed vacuum system
described by the Hamiltonian \eqref{hamSO3vacmod}.

In particular, the expression \eqref{mass} can be used to remove the radial
derivative that appears on the right-hand side of the bracket between two Hamiltonian
constraints and write it in terms of the mass and the three free functions
$f$, $g$ and $v$ as follows:
\begin{align}\label{hhmass}
   \big\lbrace H[\lapse_1],H[\lapse_2]\big\rbrace &= D\left[\mathcal{N}\frac{{E^r}}{{E^\varphi}^2}\frac{g^2}{2}\left(\frac{\partial }{\partial K_\varphi}\left[\frac{1}{g}\frac{\partial f^2}{\partial K_\varphi}\right]-\frac{1}{g}\frac{\partial^2 \ln{g}}{\partial K_\varphi^2}\left(1+f^2-\frac{2m}{v}\right)\right)\right].
\end{align}
In principle, this expression has been derived only for the vacuum case. But note that, whenever the model coupled
to matter admits a vacuum limit, the specific form of the functions $f$, $g$ and $v$ could be read off from the
corresponding vacuum Hamiltonian constraint and the mass would be the scalar quantity defined by the
expression \eqref{mass}.
In this way, the deformed algebra could also be expressed in this last form,
even under the presence of matter. 
Particularly, we will make use of this fact for the polymeric model that will be presented in the next section.

\section{Polymeric deformations}\label{polymerization}

In this section, we will focus on constructing a deformed Hamiltonian with
corrections inspired by loop quantum gravity, in particular
with the so-called holonomy modifications. In effective models for
cosmological scenarios, these corrections are
responsible for the quantum bounce that replaces the initial singularity.
Holonomy corrections are usually implemented by a polymerization procedure,
which lies in replacing connection components by periodic and bounded functions.
Note that, in our setup, the polymerization of ${K_r}$ and ${P_\phi}$ is completely
ruled out as the dependence of the modified Hamiltonian \eqref{hamalfa0} on these
momenta is completely fixed. Therefore, our goal will be to explore the possibility of considering
free functions of ${K_\varphi}$ in \eqref{hamalfa0} that satisfy the consistency equations \eqref{consistencyeqs}.

\subsection{An effective Hamiltonian with holonomy corrections}\label{apolymerham}

We will try to construct the effective Hamiltonian in such a way that it remains as close as possible
to its classical form. More precisely, we will choose
the classical form for the correction functions related to the matter Hamiltonian, that is,
$f_{3}=-1/{E^r}$ and $f_{4}=-{E^r}$ { (though the global factor $g$ might still couple holonomy corrections to the matter variables).}
In addition, the dependence on the variable $\phi$
will be assumed to appear only in a classical potential term $V(\phi)$; that is,
the function $f_0$ in \eqref{hamalfa0} will be decomposed as
$f_0(E^r,K_\varphi,\phi):=\widetilde{f}_0(E^r,K_\varphi)-2 V(\phi)/g(E^r,K_\varphi)$
\footnote{One could choose the potential as a function of the radial triad $E^r$, that is $V=V(E^r,\phi)$,
and \eqref{eq.polsolution} would still be a solution of the consistency equations.}. 
Under these assumptions, it is possible to obtain the general solution to the system
\eqref{consistencyeqs}:

\vspace{-12pt}
\begin{multicols}{2}
\begin{subequations}\label{eq.polsolution}
\begin{align}
f_{0}&=\frac{2}{g}\left(\frac{\partial {f^2}}{\partial E^r}+\left(1+f^2\right)\frac{\partial\ln v}{\partial E^r} -V(\phi)\right) \label{eq.f0pol},\\
 f_1&=\frac{2}{g}\frac{\partial {f^2}}{\partial K_\varphi},\\
  f_2&=-\frac{1}{2}\frac{\partial \ln{(g v)}}{\partial E^r},
\end{align}
\columnbreak
\begin{align}
&\nonumber\\
  h&=h_{4}=-\frac{1}{2}\frac{\partial \ln{g}}{\partial K_\varphi},\\
h_{0}&=h_{1}=h_{2}=h_{3}=0,\\
g&=\left(\frac{\partial f}{\partial K_\varphi}\right)^{\!2}\left(1+\frac{w}{f^2}\right)^{\!-1},\label{eq.globalfactorpol}
\end{align}
\end{subequations}
\end{multicols}
\vspace{-12pt}
\noindent
in terms of the integration functions $f=f({E^r},{K_\varphi})$, $v=v(E^r)$ and $w=w(E^r)$. 
Note that we have written this solution in a very similar form to the one found in Sec.~\ref{vacuum} for vacuum.
In particular, up to
the potential term $V(\phi)$, the functions $f_0$, $f_1$, $f_2$ and $h$ take exactly the same form
as in the vacuum case. 
The main difference
is that, whereas in the vacuum model the global factor $g$ was completely free, in this case $g$
is given in terms of $f$ and $w$ by relation \eqref{eq.globalfactorpol}. In essence, the freedom
in the deformation functions has been reduced: a free function of two variables
$g=g(E^r,K_\varphi)$ in the vacuum case has been replaced by a free function of just one variable $w=w(E^r)$
in the model coupled to matter.

Classically, the free function $f$ is given by $f(E^r,K_\varphi)= K_\varphi$. Therefore, in order to recover
the classical Hamiltonian, this imposes the classical limits $v\rightarrow\sqrt{E^r}$ and $w\rightarrow0$
for the other two free functions.
As we are looking for the polymerized Hamiltonian that is closest to its classical form,
we fix these last two functions to their classical values\footnote{{  Nonetheless,
note that any functions $v(E^r)$ and $w(E^r)$, such that $v\rightarrow\sqrt{E^r}$ and $w\rightarrow0$ when $f(E^r,K_\varphi)\rightarrow K_\varphi$, would also produce an anomaly-free constraint with the correct classical limit.}}.
In this way, gathering together all the above results, we obtain the deformed Hamiltonian constraint,
\begin{align}\label{polhamimproved}
\ham &=  -\frac{{E^\varphi}}{2\sqrt{{E^r}}}\left( 1+ f^2 +4{E^r}f\frac{\partial f}{\partial {E^r}}\right) -2\sqrt{{E^r}}{K_r}f \frac{\partial f}{\partial {K_\varphi}}
+\left(\frac{({E^r}')^2}{8\sqrt{{E^r}}{E^\varphi}}-\frac{\sqrt{{E^r}}}{2{E^\varphi}^2}{E^r}'{E^\varphi}'  +\frac{\sqrt{{E^r}}{E^r}''}{2{E^\varphi}}\right)\left(\frac{\partial f}{\partial {K_\varphi}}\right)^{\!2}\nonumber\\
&+\left(\frac{P_\phi^2}{2\sqrt{{E^r}}{E^\varphi}}+\frac{{E^r}^{3/2}({\phi}')^2}{2{E^\varphi}}\right)\left(\frac{\partial f}{\partial {K_\varphi}}\right)^{\!2}+\sqrt{{E^r}}{E^\varphi} V({\phi})
+\frac{\sqrt{{E^r}}}{2{E^\varphi}^2}{E^r}'\big({E^r}'{K_r}+{\phi}'{P_\phi}\big)\frac{\partial f}{\partial {K_\varphi}}\frac{\partial^2 f}{\partial K_\varphi^2}
\nonumber\\
&+\frac{\sqrt{{E^r}}}{2{E^\varphi}}({E^r}')^2\frac{\partial f}{\partial {K_\varphi}}\frac{\partial^2 f}{\partial {E^r} \partial {K_\varphi}},
\end{align}
which, for any function $f=f(E^r, K_\varphi)$, forms a first-class algebra along with the classical diffeomorphism
constraint. The corresponding Poisson brackets will be given by \eqref{hh}, where one should replace the form \eqref{eq.polsolution} for the functions $g$, $f_1$ and $h$. 
As already commented above, the particular choice $f(E^r,K_\varphi)={K_\varphi}$ reproduces
the classical Hamiltonian. Moreover, the free dependence of $f$ on $E^r$ allows us to
introduce a scale-dependence in the holonomy corrections. Nonetheless, as noted for similar models
in previous studies
\cite{tibrewala}, this freedom spoils the periodicity in $K_\varphi$, at least for simple sinusoidal
choices of the free function $f$. For instance, if one considers $f=\sin(\mu(E^r)K_\varphi)/\mu(E^r)$ the derivatives of this function with respect
to $E^r$ that appear in the last expression would introduce a non-periodic dependence on $K_\varphi$
in the Hamiltonian.

Therefore, in order to obtain a $K_\varphi$-periodic Hamiltonian constraint, let us restrict ourselves
to the case with a scale-invariant function $f=f(K_\varphi)$. Furthermore,
as we look for bounded functions, we can set $f=\sin{(\lambda {K_\varphi})}/\lambda$,
with a real constant $\lambda$, which provides the classical limit for $\lambda\rightarrow 0$
and yields the polymeric constraint,
\begin{align}\label{polham}
\ham^{({pol})} &=  -\frac{{E^\varphi}}{2\sqrt{{E^r}}}\left(1+\frac{\sin^2{{(\lambda {K_\varphi})}}}{{{\lambda^2}}}\right)  -\sqrt{{E^r}}{K_r}\frac{\sin{(2\lambda {K_\varphi})}}{\lambda}\left(1+\left(\frac{\lambda {E^r}'}{2{E^\varphi}}\right)^{\!2}\right) \nonumber\\
&+\left(\frac{({E^r}')^2}{8\sqrt{{E^r}}{E^\varphi}}  -\frac{\sqrt{{E^r}}}{2{E^\varphi}^2}{E^r}'{E^\varphi}'
 +\frac{\sqrt{{E^r}}}{2{E^\varphi}}{E^r}'' +\frac{P_\phi^2}{2\sqrt{{E^r}}{E^\varphi}}+\frac{{E^r}^{3/2}({\phi}')^2}{2{E^\varphi}}\right)\cos^2{(\lambda {K_\varphi})} \nonumber\\
&+\sqrt{{E^r}}{E^\varphi}{V}({\phi}) -\frac{\sqrt{{E^r}}}{4{E^\varphi}^2}{E^r}'{\phi}'{P_\phi}\lambda\sin{(2\lambda {K_\varphi})},
\end{align}
which obeys the deformed Poisson bracket:
\begin{align}\label{hhpol}
   \big\lbrace H^{({pol})}[\lapse_1],H^{({pol})}[\lapse_2]\big\rbrace &= D\left[\beta{{E^r}}{{E^\varphi}^{-2}}\left(\lapse_1\lapse_2'-\lapse_1'\lapse_2\right)\right],
\end{align}
with {\small{$H^{({pol})}[N]:=\int {\rm d}r N {\cal H}^{({pol})}$}} and the deformation
function,
\begin{align}\label{deformationk}
    \beta := \cos^2{(\lambda {K_\varphi})}\left( 1+\left(\frac{\lambda {E^r}'}{2{E^\varphi}}\right)^{\!2}\right) .
\end{align}
The constraint \eqref{polham} is regular for all possible values of $K_\varphi$. Concerning dynamical properties, this Hamiltonian is periodic in $\lambda K_\varphi$
with period $\pi$. Hence, further dynamical studies can be restricted to the range $\lambda K_\varphi\in [0,\pi)$.
In this range, the value $\lambda K_\varphi=\pi/2$, where the deformation function $\beta$ vanishes,
corresponds to a symmetry point. 
The Hamiltonian is invariant under reflection of $\lambda K_\varphi$ around that point $\lambda K_\varphi\rightarrow \pi-\lambda K_\varphi $, and the
change of sign of the other momenta $K_r\rightarrow - K_r$
and $P_\phi\rightarrow-P_\phi$, which reflects a time-symmetry of the model. Regarding the equations of motion, they are also invariant under the commented transformation
as long as one also considers the inversion of time $t\rightarrow -t$. As will
be explained in the next section, the present model is closely related to the one presented in \cite{bgp2},
and the same physical conclusions have been reached in \cite{Bojowald:2021isp}: the surface where the deformation function
$\beta$ vanishes can be understood as a time-reversal surface. However, in \cite{Bojowald:2021isp} it is argued that
this model does not respect covariance since, $K_\varphi$ not being a spacetime scalar,
the surface $\beta=0$ is not covariantly defined. Nonetheless, we will show
that in the present model the deformation function (and, particularly, its zeros)
can indeed be covariantly represented by making use of the mass function \eqref{mass} defined above.

More precisely, following the derivation performed in Sec.~\ref{vacuum}, we can define the mass for this
polymerized Hamiltonian as follows,
\begin{align}\label{eq.masspol}
    m_{{pol}}&:=\frac{\sqrt{{{E^r}}}}{2}\bigg(1+\frac{\sin^2{(\lambda {K_\varphi})}}{\lambda^2}-\left(\frac{{E^r}'}{2{E^\varphi}}\right)^{\!2}\cos^2{(\lambda {K_\varphi})}\bigg).
\end{align}
In vacuum, this will be a weak observable and thus it will be conserved through evolution,
in contrast to the model coupled to matter where it is a dynamical quantity. In either case,
the deformation of the structure function can be written in terms of this ``polymerized mass'':
\begin{align}\label{deformationm}
    \beta = 1 +\lambda^2\left(1-\frac{2m_{{pol}}}{\sqrt{E^r}}\right) .
\end{align}
In this way, $\beta$ is completely determined by scalar functions. Since, by the definition \eqref{deformationk},
the deformation function is non-negative, this last relation provides a lower bound for
the radial component of the triad:
\begin{align}\label{minq1}
    \sqrt{E^r} \geq\frac{2\lambda^2m_{{pol}}}{1+\lambda^2}, 
\end{align}
where the saturation of this bound corresponds to the point where $\beta=0$.
In vacuum $m_{pol}$ represents a generalized Schwarzschild mass, which should be
positive, and thus this result ensures that the system can not reach the classical singularity
at $E^r=0$. When considering a matter field, $m_{pol}$ will not be an observable
and it will have a non-trivial evolution.
Therefore, in order to check whether the system can reach the classical singularity,
one should numerically solve the equations of motion and, in particular,
study how $m_{pol}$ scales 
as $E^r\rightarrow 0$.

\subsection{Back to general relativity through a canonical transformation}\label{backtoGR}

The first terms in the Hamiltonian \eqref{polham}, that go with a sine function
of the curvature component $K_\varphi$,
are the usual correction terms considered in different polymeric models in vacuum.
However, these terms by themselves do not remain covariant under the presence of matter with local degrees of freedom.
In order to ensure the covariance of the model, our derivation has led us to include two
additional corrections: the cosine function of the curvature component that
multiplies several terms in \eqref{polham}, as well as the coupling between radial derivatives
of matter and geometric variables that appears in the last term of \eqref{polham}.
In order to understand better these new contributions, we will relate our derivation
to the recent proposal in \cite{bgp2}, where a polymerized constraint is obtained by means of
a canonical transformation of the classical phase-space variables.

Let us consider the following canonical transformation on the classical model,
as the generalization of the one proposed in \cite{bgp2}:
\begin{align}\label{cantrans}
    {K_\varphi}\rightarrow  f({E^r},{K_\varphi}),
    \quad {E^\varphi}\rightarrow  {{E^\varphi}} \left(\frac{\partial f}{\partial {K_\varphi}}\right)^{\!-1},\quad{E^r}\rightarrow {E^r},\quad {K_r}\rightarrow {K_r} +{E^\varphi}\frac{\partial f}{\partial {E^r}}\left(\frac{\partial f}{\partial {K_\varphi}}\right)^{\!-1}.
\end{align}
The transformed Hamiltonian $\widetilde {\cal H}$ is straightforwardly obtained just by considering the
above transformation in the classical constraint ${\cal H}^{(class)}$ \eqref{hamspherclass}. Concerning the
Poisson algebra, the bracket between the transformed Hamiltonian $\widetilde{\cal H}$
and diffeormorphism constraint ${\cal D}$ is unchanged
but, due to the presence of phase-space variables on the right-hand side of \eqref{hh},
the bracket between two transformed Hamiltonian constraints acquires a deformation,
\begin{equation}
 \{\widetilde{H}[N_1],\widetilde{H}[N_2]\}=D\left[{\cal N}\frac{E^r}{(E^\varphi)^{2}}\left(\frac{\partial f}{\partial K_\varphi}\right)^{\!2}\right],
\end{equation}
where $\widetilde H$ stands for the smeared form of $\widetilde{\cal H}$.
In fact, since the classical Hamiltonian is linear in $E^\varphi$ and $K_r$,
the new constraint $\widetilde{\cal H}$ would be divergent in the zeros
of the deformation function $\partial f/\partial {K_\varphi}$. One could try
to get rid of those potential infinities through a rescaling
$\widetilde{{\cal H}}':=\widetilde{\ham}\,\partial f/\partial {K_\varphi}$,
but the constraint $\widetilde{\cal H}'$ would not be the canonical generator of infinitesimal
transformations since the bracket with itself would be schematically given by,
$$\{\widetilde{H}',\widetilde{H}'\}=\widetilde{H}'+D,$$ $\widetilde{H}'$ being
the smeared form of $\widetilde{\cal H}'$.
Nonetheless, it can be shown that the following combination with the diffeomorphism constraint,
\begin{align}
    \ham = \left(\widetilde{\ham} -\frac{\sqrt{{E^r}}{E^r}'}{2{E^\varphi}^2}\frac{\partial^2 f}{\partial K_\varphi^2}\diff\right)\frac{\partial f}{\partial {K_\varphi}},
\end{align}
produces the canonical result for the bracket $\{H,H\} = D$. In fact, this last Hamiltonian turns out to be
exactly the same as the one we have obtained above \eqref{polhamimproved}. Therefore,
the extra couplings ${K_r}{E^r}'$ and ${P_\phi}{\phi}'$ that appear in that constraint
can be understood as coming from the linear combination with the diffeomorphism constraint
performed in the last expression.

Hence, we have provided a way to reach a polymerized constraint \eqref{polhamimproved} which, outside the surface where the
deformation function vanishes, can be matched
to the one constructed in \cite{bgp2}\footnote{In that reference, $f=\sin{(\lambda K_\varphi)}$ and there is an additional transformation for the scalar field $({\phi},{P_\phi})$ which we did not consider here.}. Nonetheless, our constraint \eqref{polhamimproved}
is regular for all possible values of $K_\varphi$, and one
avoids the divergences arising from the canonical transformation above, as well
as potential ill-defined lapse and shift rescalings at the zeros of the deformation function.

Finally, let us point out that, since the Hamiltonian presented in \cite{bgp2}
contains derivatives of the momenta ($K_r, K_\varphi$, $P_\phi)$, it is not included in
the family of deformed Hamiltonians \eqref{hamalfa0}, which has been constructed by
assuming the ansatz \eqref{ham}. That is why, in order to relate both approaches,
it is necessary to consider the above linear combination with the
diffeomorphism constraint.

\section{Discussion}\label{conclusion}

In this article, we have performed a comprehensive study of Dirac's deformation algebra for a general constraint with quadratic dependence on radial derivatives. Starting from the ansatz \eqref{ham}, this approach has allowed us to
systematically construct modified spherically symmetric models coupled to a scalar matter field.
The main result of our study is encoded in the family of modified Hamiltonian constraints \eqref{hamalfa0},
along with the consistency equations \eqref{consistencyeqs}. Any Hamiltonian of this form
respects covariance inasmuch as it forms a first-class constraint system alongside with the
classical diffeomorphism constraint. In particular, concerning the dependence on momenta $(K_r,K_\varphi, P_\phi)$,
it can be seen that, demanding the closure of Dirac's commutation relations, 
explicitly fixes the dependence of the Hamiltonian \eqref{hamspherclass} on the momenta $(K_r,P_\phi)$,
whereas the consistency equations severely restrict the allowed dependence on ${K_\varphi}$.

Although the full consistency-equation system is highly coupled, we have been able to provide
the general solution for the vacuum case. The resulting vacuum Hamiltonian \eqref{hamSO3vacmod} has two free
functions, which depend on $E^r$ and $K_\varphi$, and an additional freedom regarding only $E^r$. 
Moreover, we have found a weak Dirac observable \eqref{mass} that can be identified
with the mass of the modified system, and its classical limit reproduces the usual
Schwarzschild mass.

In the last section, under certain assumptions for some of the free functions coupled to the matter sector,
we have solved the consistency equations for the case coupled to matter
in order to obtain a modified Hamiltonian with some free dependence of
the curvature component ${K_\varphi}$, which could be understood as a covariant polymerization of the system. This solution has led us to the constraint \eqref{polhamimproved}, which is parametrized by
one free function of $E^r$ and ${K_\varphi}$.
Therefore, this analysis opens the possibility to study (scale-dependent) holonomy corrections
for this model coupled to matter with local degrees of freedom.
{  Nonetheless, let us stress that probably the obtained solution is not unique,
and that the complexity of the system \eqref{consistencyeqs} might leave room for other
different polymerized constraints.}

Finally, by choosing a specific form for the mentioned function, we have obtained the polymeric constraint \eqref{polham},
which is regular for all values of $K_\varphi$ and forms a first-class algebra with the brackets \eqref{hhpol}
corrected with the deformation function $\beta$.
This result can be seen as a family of effective midisuperspaces depending on a free real parameter $\lambda$,
with the specific case $\lambda=0$ corresponding to general relativity. In fact, as has been shown in the
last subsection, these models can be mapped to general relativity through a (divergent) canonical transformation,
similar to the one considered in \cite{bgp2}, and a rescaling of the lapse and shift. In consequence, the effective midisuperspaces are locally equivalent to the classical theory \cite{bgp2,Bojowald:2021isp}, except in the surfaces of vanishing deformation function $\beta$, which can be interpreted as time-reversal surfaces.
An issue raised in \cite{Bojowald:2021isp} is that the definition of these surfaces breaks
the covariance of the system as they are defined in terms of the values of $K_\varphi$,
which is not a spacetime scalar. Nonetheless, in our model we have been able to write the
deformation function in terms of the mass of the system \eqref{eq.masspol}
and the radial component of the triad $E^r$. This has shown that, at least in vacuum,
the commented surfaces correspond to a minimum value of $E^r$ and thus the classical
singularity is unreachable by the modified system.

\vspace{0.4cm}
\begin{center}
    \textbf{Acknowledgements}
\end{center}
\vspace{8pt}
\noindent {AAB acknowledges financial support from the FPI fellowship \mbox{PRE2018-086516} of the Spanish Ministry of Science, Innovation and Universities. This work is funded by Project \mbox{FIS2017-85076-P} (MINECO/AEI/FEDER, UE) and Basque Government Grant \mbox{No.~IT956-16}.}

\bibliographystyle{ieeetr}

\end{document}